\documentclass[11pt,onecolumn]{article}
\usepackage{ISG_style}
\usepackage{times}
\usepackage{graphicx,subfigure}
\usepackage{color}
\usepackage{amsmath,amssymb} 
\usepackage{dsfont}
\usepackage{tabularray}
\usepackage{cite}
\usepackage{hyperref}
\usepackage{epstopdf}
\usepackage{url}
\usepackage{flushend}

\usepackage{graphicx}
\usepackage{multirow}
\usepackage{enumitem}
\usepackage[table]{xcolor}

\usepackage{algorithm}
\usepackage{makecell}
\usepackage{algcompatible}
\usepackage{amssymb, amsmath,array, amscd, bm}
\usepackage{algpseudocode}
\usepackage{mathtools}
\usepackage{verbatim}

\usepackage{tikz}
\usetikzlibrary{shapes.geometric, arrows}
\usepackage[]{todo notes}
\usepackage{nicefrac}

\usepackage[font=small,justification=centering]{caption}
\ifpdf
    \graphicspath{{figures/PNG/}{figures/PDF/}{figures/}}
\else
    \graphicspath{{figures/EPS/}{figures/}}
\fi

\makeatletter
\renewcommand{\Function}[2]{%
  \csname ALG@cmd@\ALG@L @Function\endcsname{#1}{#2}%
  \def\jayden@currentfunction{#1}%
}
\newcommand{\funclabel}[1]{%
  \@bsphack
  \protected@write\@auxout{}{%
    \string\newlabel{#1}{{\jayden@currentfunction}{\thepage}}%
  }%
  \@esphack
}
\makeatother

\newcommand*{\boldone}{\text{\usefont{U}{bbold}{m}{n}1}}

\newcommand{\Precision}{{\sf precision}}
\newcommand{\Regret}{{\sf regret}}

\DeclareMathOperator*{\argmin}{arg\,min}  
\DeclareMathOperator*{\argmax}{arg\;max}

\newcommand{\cpath}{{\sf C}\text{-}{\sf Path}}
\newcommand{\lingam}{{\sf LiNGAM}}
\newcommand{\ica}{{\sf ICA}}
\newcommand{\pa}{{\rm pa}}
\newcommand{\ch}{{\rm ch}}
\newcommand{\tv}{{\sf cost}}
\newcommand{\baseload}{{\sf base\_load}}
\newcommand{\curly}{\mathrel{\leadsto}}

\tikzstyle{startstop} = [rectangle, rounded corners, 
minimum width=2cm, 
minimum height=0.75cm,
text centered, 
draw=black, 
fill=yellow!10]

\tikzstyle{io} = [trapezium, 
trapezium stretches=true, 
trapezium left angle=70, 
trapezium right angle=110, 
minimum width=2cm, 
minimum height=0.75cm, text centered, 
draw=black, fill=blue!30]

\tikzstyle{process} = [rectangle, 
minimum width=2.5cm, 
minimum height=0.75cm, 
text centered, 
text width=2cm, 
draw=black, 
fill=gray!10]

\tikzstyle{decision} = [diamond, 
aspect=2, 
text centered, 
draw=black, 
fill=blue!10]
\tikzstyle{arrow} = [thick,->,>=stealth]

\title{\bf \huge Cascading~\textcolor{black}{Failure} Prediction via Causal Inference}

\date{}

\author{Shiuli~Subhra~Ghosh\thanks{Shiuli Subhra Ghosh, Anmol Dwivedi, and Ali Tajer are with Rensselaer Polytechnic Institute, Troy, NY.} \qquad Anmol~Dwivedi \footnotemark[1] \qquad Ali~Tajer \footnotemark[1]
\qquad Kyongmin~Yeo \thanks{Kyongmin~Yeo and Wesley~M.~Gifford are with IBM T.J. Watson Research Center, Yorktown, NY.}\qquad Wesley~M.~Gifford \footnotemark[2]
}

\begin{document}

\maketitle
\allowdisplaybreaks

\begin{abstract}

Causal inference provides an analytical framework to identify and quantify cause-and-effect relationships among a network of interacting agents. This paper offers a novel framework for analyzing cascading \textcolor{black}{failures} in power transmission networks. This framework generates a \emph{directed} latent graph in which the nodes represent the transmission lines and the directed edges encode the cause-effect relationships. This graph has a structure distinct from the system's topology, signifying the intricate fact that both local and non-local interdependencies exist among transmission lines, which are more general than only the local interdependencies that topological graphs can present. This paper formalizes a causal inference framework for predicting how an emerging anomaly propagates throughout the system. Using this framework, two algorithms are designed, providing an analytical framework to identify the most likely and most costly cascading scenarios. The framework's effectiveness is evaluated compared to the pertinent literature on the IEEE $14$-bus\textcolor{black}{, $39$-bus, and $118$-bus systems}.
\end{abstract}

\section{Introduction}
\label{sec:Introduction}

Causal reasoning is an analytical approach to understanding the cause-and-effect relationships among the events in a network of agents with complex interactions. It involves identifying how one event or variable influences or leads to another event or variable. Specifically, in complex networks of interconnected nodes, causal learning provides a theoretically principled framework to uncover and quantify how changing the behavior of one network component affects others. 

In this paper, we formalize a causal framework for predicting the potential cascading effects arising from \textcolor{black}{line failures} in power transmission networks. The framework leverages the intricate interconnectivity among the components (generators, buses, transmission lines, and loads). This intricate interconnectivity serves as both a resilience measure for transmission networks, incorporating redundancies in infrastructure and resources, and a source of complex and unforeseen combinations of \textcolor{black}{line failures}. Before specifying our approach, we review the literature closely relevant to this paper's scope.

\vspace{.05 in}
\noindent \textbf{Scenario-based simulations:} 
Scenario-based simulations are effective for generating failure scenarios and evaluating their consequences. Such approaches assess the risks and likelihood associated with diverse scenarios and are effective for risk and contingency analysis~\cite{OPA_4, OPA_3_improved, OPA_5_quasi_dynamic}. Specifically, these methods employ numerical simulations that iteratively solve power flows, thereby determining the sequence of quasi-static transmission line overflows~\cite{OPA_1, OPA_6_initial_model}. DC optimal power flow (OPF) analysis provides a computationally efficient approach to scenario generation, but it lacks sufficient versatility to represent all possible cascading failure scenarios~\cite{OPA_2}. To achieve higher versatility, adopting AC-OPF~\cite{OPA_7} further leads to better scenario representation and predictive power at the expense of higher computational complexity. Such complexity grows rapidly as the system's scale and forecasting horizons grow. Finally, since it is impossible to create all conceivable scenarios, these methods tend to focus on the more probable scenarios, potentially overlooking low-probability yet high-risk situations.

\vspace{.05 in}
\noindent \textbf{Topology-guided approaches:} To overcome the limitations of scenario-based simulations, topology-based approaches have been introduced to complement or substitute simulations, capitalizing on the legacy topology models. By analyzing the power grid's topological structure, these methods estimate the probable pathways of cascading \textcolor{black}{failures}~\cite{ps_top_italian, ps_top_structural_vul,J33}. These methods implicitly assume that \textcolor{black}{transmission line failures propagate through paths connecting the components. This indicates that a transmission line failing causes failures in its adjacent transmission lines.} In these methods, topological metrics (e.g., degree distribution and betweenness centrality) are effective for predicting the propagation of failures throughout the network~\cite{ps_top_1, ps_top_2}. Topology-guided approaches, however, do not incorporate electrical quantities or power flow constraints into the modeling of cascading failure, as highlighted in~\cite{ps_top_hines, Fault_chain_theory}. Recent advancements in this domain have incorporated electrical constraints along with the grid topology by leveraging efficient deep learning techniques to predict failure cascade~\cite{ML_2} and identify high-risk cascades~\cite{ML_1, grnn}. \textcolor{black}{While providing effective predictive power, such black-box approaches are not amenable to interpreting the root causes of failures. Consequently, their predictive power is limited by the data on which they are trained. Hence, unlikely but impactful failure scenarios that are not well-represented in the training data can be mis-predicted.}

\vspace{.05 in}
\noindent \textbf{Latent space inference:} 
In addition to the topology, other relevant network variables (e.g., state parameters) can track the likelihood of failure sequences when predicting \textcolor{black}{cascades}. This has been the premise for designing \emph{influence graphs}~\cite{Hines_first_non_top, branching_process, IG_1, IG_2, IG_3} and~\emph{interaction} graphs~\cite{Interaction_graph_1, Interaction_graph_2, Interaction_graph_4, Interaction_graph_5}, which recognize the existence of a latent process beyond topology that plays a crucial role in predicting failure sequences. The objectives of these models are to infer the latent structure influencing \textcolor{black}{failure} propagation and to integrate this with topological information to enhance predictive accuracy. Specifically, influence graphs encode the conditional probability that one component will fail, given the failure of another component, assuming a one-step Markovian transition. This implies that a \textcolor{black}{failed} component uniquely determines the probabilities of subsequent failures. At the expense of being a simplifying statistical assumption, this Markovian premise facilitates computationally efficient solutions.

\vspace{.05 in}
\noindent \textbf{Causal graphs:} In this paper, we construct a graphical model in which the nodes represent the network components, and the edges collectively encode the complex cause-effect relationships among the network components, \cite{koller2009probabilistic, Pearl_book}. \textcolor{black}{Such graphs are referred to as {\em causal graphs}} and aim to address the following question: if we fix the behavior of all the components in the graph and vary the behavior of only one component, how does that one change affect the behavior of the rest of the components? The answer to this question specifies two pieces of information for the causal graph: (i)~the cause-effect relationships among the nodes, which are depicted by \emph{directed} edges, and (ii)~the extent (power) of cause-effect relationships, which are specified by \emph{structural equation models}~\cite{RESIT, ICA_LiNGAM, Direct_LiNGAM, DAG_NOTEARS}. \textcolor{black}{Such a latent causal graph will share some similarities with the network's connectivity topology but will also reveal significant differences. These differences indicate how failures propagate along paths that do not conform to the system's topology, ultimately offering greater predictive power in determining the consequences of an anomaly.}

\textcolor{black}{There are several key differences between causal inference and machine learning (ML)--based solutions. To lay the context, the most accurate way to capture all the delicate interactions among the variables involved in a power system is through a set of coupled equations. These equations precisely specify how changing one variable influences others in the systems. Since accurately characterizing such equations can be prohibitive, ML-based solutions have been used as alternatives. These solutions, however, capture only statistical associations (e.g., correlation) and cannot predict the influence of varying one variable on the others. Causal inference fills the gap between a coupled set of equations and ML-based solutions. They are data-driven (similar to ML-based solutions), and they also can predict the mutual influence of variables (similar to the equations)}.

\textcolor{black}{Hence, the causal inference framework complements ML-based solutions in three ways. First, the predictive power of ML-based models is limited to the scenarios represented in their training data, whereas causal inference learns the cause-effect relationships among variables, enabling accurate predictions even in previously unseen scenarios. Second, causal inference provides interpretable predictions, explaining not only how cascading failures propagate but also why they occur, unlike the black-box nature of ML models. Lastly, causal models can be used for scenario generation by varying inputs, offering an alternative to costly simulations typically required for training ML algorithms.}

\vspace{.05 in}
The causal framework is \emph{data-driven}, can complement the existing methods, and has the following three features that distinguish it from the three broad approaches discussed earlier: (i) it is computationally efficient and does not face the prohibitive complexity that scenario-based simulations face; (ii) it accounts for \textcolor{black}{traceable} non-local propagation of cascades, which \textcolor{black}{ML-based} topology-guided approaches do not account for; and (iii) it leverages the data structure embedded in a latent space to uncover cause-effect relationships, exceeding beyond the existing approaches that focus solely on statistical relationships. 

\noindent \textcolor{black}{\textbf{Contributions:} 
The contributions of this paper can be summarized as follows. 
\begin{enumerate}
    \item We propose a causal model that uses training data to learn a \emph{latent} causal graph that captures the cause-effect relationships among the failures of transmission lines from observational power flow data. This facilitates transcending only learning statistical associations provided by ML. 
    \item To effectively use the anomaly-free causal graph learned, we design \emph{intervention} mechanisms to model failures. This facilitates recovering the causal model of the system after failures by a computationally efficient update to the anomaly-free causal model of the system and prevents re-learning a causal model for every cascading scenario. 
    \item We design two algorithms using causal path analysis, which provide for real-time prediction of cascading failure as a failure emerges and for identifying critical cascading scenarios. 
    \item We provide performance evaluations through extensive experiments on the IEEE $14$-bus, $39$-bus, and $118$-bus systems and compare the results with those of the most relevant existing approaches based on influence graphs~\cite{IG_1} and graph neural networks~\cite{ML_2}.
\end{enumerate}
    }


\section{Problem Formulation}
\label{sec: Problem Formulation}

\subsection{\textcolor{black}{Failure} Model}
\label{sec: Anomaly Model}

Consider a power transmission network consisting of $N$ transmission lines and $L$ loads. We denote the absolute active power flow in line $i \in  [N] \dff \{1, \dots, N\}$ at the discrete time instant $t \in \N$ by $P_i[t]$ (in MWs). Accordingly, we define 
\begin{align}
    \label{eq: flows}
        \bP[t]\dff \left[ P_1[t],\dots, P_N[t]\;\right]^\top\ . 
\end{align}
Each line $i \in [N]$ is subject to maximum power flow constraint denoted as $P_i^{\sf max}$. Various internal factors, such as system instabilities, or external factors, such as weather conditions, can potentially contribute to \textcolor{black}{transmission line failures}. 
To model the presence and extent of potential \textcolor{black}{failures} in line $i \in [N]$, we define $S_i[t]$ as an anomaly quantification index that captures the deviation in absolute power flow in line $i$ at time $t$ from the flow at time $t-1$ given by
\begin{align}
    \label{eq: anomaly_metric}
     S_i[t] \dff \frac{P_{i}[t]-P_{i}[t-1]}{P_i^{\sf max}} = \frac{\Delta P_i[t]}{P_i^{\sf max}} \ .
\end{align}
In order to account for the differences in maximum capacities across transmission lines, we normalize such flow deviations $\Delta P_i[t]$ by their respective maximum capacities, $P_i^{\sf max}$. Accordingly, the anomaly vector at time $t$ associated with all the lines is denoted by $\bS[t]\dff [S_1[t],\dots, S_N[t]]^\top$. 
To effectively represent the severity of \textcolor{black}{a failure}, we discretize $S_i[t]$ into $T$ distinct levels, thereby assigning different \textcolor{black}{failure} states $\tilde S_{i}[t]\in \mcS\dff \{s_1,\dots, s_T\}$ to represent varying degrees of anomalies. For instance, when we are concerned about only distinguishing line outages, we set $T=2$, based on which $\tilde S_i[t]=s_1$ indicates that line $i$ is healthy, while $\tilde S_i[t]=s_2$ signifies that the line is in an outage. Similarly, when failures are specified as significant deviations of power flow from the expected ranges, the set $\{s_2,\dots,s_T\}$ can be specified to represent a desired level of granularity in these deviations. 
Consequently, we define the discrete \textcolor{black}{failure} vector at time $t$ by $\tilde{\bS}[t] \dff [\tilde{S}_1[t], \dots, \tilde{S}_N[t]]$. Our focus is on persistent \textcolor{black}{failures}, that is once a line is \textcolor{black}{failed} it remains in the same state until a remedial action is exerted. Finally, we define the set $\mcU_t$ to specify the set of transmission lines that \textcolor{black}{fail} at time $t$ by
\begin{align}
    \label{eq: anomalous set}
        \mcU_t \dff \{i \in [N]: \tilde S_i[t] \neq 1, \tilde S_i[t-1] = 1 \}\ . 
\end{align}

\subsection{Cascading Cost}
\label{sec: Cascading Cost}

When an emerging \textcolor{black}{failure} in the system is severe enough, it can stress the system, e.g., a line outage can lead to overloads. These, potentially, can lead to additional \textcolor{black}{failure} events, leading to a sequence of \textcolor{black}{line failures} (cascade). 
To formalize a model for cascading \textcolor{black}{failures} and their associated risks, we start by specifying \textcolor{black}{a failure}-free system that precedes a cascade and denote the associated power flow and anomaly \textcolor{black}{vector} before \textcolor{black}{a failure} starts emerging by $\bP[0]$ and $\bS[0]$, respectively. We denote the number of stages over which the sequence of \textcolor{black}{failures} materializes by $M$.
The set of \textcolor{black}{line failures} that are added to the cascade in stage $m \in [M]$ is denoted by $\mcU_{m}\subseteq [N]$. Collectively, we denote the sequence of \textcolor{black}{line failures} throughout the $M$ stages by  
\begin{align}
\label{eq:sequence}
    \mcV \dff \langle \mcU_1, \dots, \mcU_{M} \rangle \ .
\end{align} Without loss of generality, we assume that at each stage only one transmission line can \textcolor{black}{fail}, i.e., $|\mcU_m| = 1$.
To capture the stress imposed on the system by \textcolor{black}{a failure} sequence, we denote the cost incurred due to cascade $\mcV$ by 
\begin{align}
\label{eq:total_vulnerability}
    \tv(\mcV) \dff \sum_{m=1}^{M}\sum_{i=1}^{N}~\textcolor{black}{\left|S_i[m]\right|}\ . 
\end{align}

\subsection{Statements of the Objectives}
\label{sec: Statements of the Objectives}

In transmission networks, cascading \textcolor{black}{failures} arise from complex, latent interactions among transmission lines, rendering the propagation unpredictable. While certain cascades may occur with high frequency, others, particularly those resulting in significant disruption, may transpire only a few times. \textcolor{black}{We denote these as \emph{most probable} and \emph{less probable} cascading sequences, respectively}. Thus, real-time prediction and identification of these \textcolor{black}{failures} are crucial to maintaining system stability and reliability. Moreover, given the varied risks associated with different cascading events due to the system's complex dynamics, it is essential for system operators to efficiently and quickly identify those cascades that result in substantial costs from flow observations and anomaly states.

In this paper, we have two objectives. The first objective pertains to the real-time prediction of \textcolor{black}{failures} that are {\bf most likely} to occur in the subsequent stage, based on observations in the past stages, and the second one focuses on the real-time prediction of the possible sequence of \textcolor{black}{failures} that are {\bf most costly}. Next, we formalize these two problems.

\vspace{.1 in}
\noindent {\bf Likely failures:} Motivated by forming fast situational awareness about the consequences of emerging \textcolor{black}{failures} at stage $m\in[M]$ based on the past states of \textcolor{black}{a failure} sequence $\mcV$, captured via $\bar \bP_{m}$ and $\bar \bS_{m}$, defined by
\begin{align}
    \bar \bP_{m} \triangleq\{\bP[i]:i\in[m]\}\ , \;\; \mbox{and} \;\;\; \bar \bS_{m} \triangleq\{\bS[i]:i\in[m]\}\ ,
\end{align}
our objective is to identify the transmission line most likely to \textcolor{black}{fail} in stage $m+1$. This is formalized as
\begin{align}\label{eq:P1}
   \mcP_1: \;  \hat \mcU_{m+1} \triangleq \argmax_{\mcU\subseteq [N] \backslash \langle \mcU_1,\dots \mcU_m \rangle, |\mcU| \leq \textcolor{black}{\lceil N \times \kappa \% \rceil }} \P (\mcU  \med \bar\bP_{m}, \bar\bS_{m})\ ,
\end{align} where $\hat \mcU_{m+1}$ represents the predicted set of potential \textcolor{black}{failures}, limited to \textcolor{black}{$\lceil N \times \kappa \% \rceil$} lines \textcolor{black}{for varying \textcolor{black}{$\kappa \%$}}, given the past network states. 

\vspace{.1 in}
\noindent {\bf Costly failures:} \textcolor{black}{Failures} often emerge gradually and can vary significantly in their impact, with some proving more costly than others. Our second objective is to develop a causal predictive model that can predict costly sequences of \textcolor{black}{failures}. To this end, we define $\mcZ$ as the set of all~\emph{dependent} \textcolor{black}{failure} sequences occurring over $M$ stages and our objective is to identify $d$ such \textcolor{black}{failure} sequences that are deemed most costly, as formalized below
\begin{align}\label{eq:P2}
    \mcP_2: \quad \{\mcV^*_{1}, \dots, \mcV^*_d\} \dff \argmax_{\{ \mcV_1, \dots, \mcV_d \}:\mcV_i\in \mcZ} \; \sum_{i = 1}^{d} \tv(\mcV_i) \ .
\end{align}
Without loss of generality, we specify these $d$ sequences such that their associated costs appear in decreasing order, i.e., $\tv(\mcV^*_1) > \tv(\mcV^*_2) > \dots > \tv(\mcV^*_{d})$. Solving $\mcP_2$ is computationally expensive as the complexity of the search space $\mcZ$ grows exponentially with the number of lines $N$ and the cascade horizon $M$, rendering solving $\mcP_2$ computationally prohibitive even for moderate values of $N$ and $M$.

\subsection{Accuracy Metrics}
\label{sec: Accuracy Metrics}

To assess the accuracy of the predictions derived from solving $\mcP_1$ and $\mcP_2$ in \eqref{eq:P1} and \eqref{eq:P2}, respectively, we present the following accuracy metrics. Corresponding to a prediction $\hat\mcU_m$ provided by $\mcP_1$ and its associated ground truth $\mcU_m$, we define a~\emph{precision} metric that quantifies the fraction of correct predictions associated with each of the $M-1$ stages within a cascading sequence. Specifically, for any cascade $\mcV$, we define
\begin{align}
\label{eq:cost-p2}
     \Precision (\mcV )\dff \frac{1}{M-1} \cdot \sum_{m = 2}^{M}  \boldone\{\mcU_{m} \in \hat \mcU_{m} \} \ ,
\end{align}
where $\boldone\{\cdot\}$ denotes the indicator function, which takes value~$1$ when the ground truth $\mcU_{m}$ lies in the predicted set $\hat \mcU_{m}$. Naturally, a higher precision value indicates higher accuracy.

We also define a~\emph{regret} metric that quantifies the accuracy of predicting the average cost associated with problem $\mcP_{2}$. In particular, consider any $d$ predicted sequences of \textcolor{black}{failures} $\bar\mcV_{d} \triangleq \{\mcV_{i}: i\in[d]\}$. We define the regret of these sequences with respect to the ground truth $\bar\mcV^{*}_{d} \triangleq \{\mcV^*_{i}: i\in[d]\}$ by
\begin{align}
\label{eq:regret}
    \Regret (\bar\mcV_{d})\dff 1- \frac{\sum_{i=1}^{d} \tv(\mcV_{i})}{\sum_{i = 1}^{d} \tv(\mcV^*_i)} \ .
\end{align}

\section{Foundations of Causal Inference}
\label{sec: Foundations of Causal Inference}

In this section, we introduce the key components of the causal inference framework, laying the groundwork for quantifying cause-effect relationships. The subsequent section will leverage this framework to model cascading \textcolor{black}{failures}, enabling real-time and accurate prediction of future \textcolor{black}{failure} sequences once \textcolor{black}{a failure} emerges in the system. 

\subsection{Formalizing Cause-Effect Relationships}
\label{subsec: Formalizing Cause-Effect Relationships}

Specifying the cause-effect relationships among a set of interacting entities (i.e., line anomaly states $S_{i}$) in an interconnected network can be formalized via two key components: (i) a directed graph that signifies the cause-effect directions among the entities, and (ii) a structural causal model (SCM) that quantifies the extent of each specific cause's effect.

\vspace{.1 in}
\noindent{\bf 1.\;Directed Graph $\mcG$:} We denote the interaction between transmission lines by a \emph{directed} graph $\mcG = ([N], \mcE)$, where the set of nodes $[N]$ represent the transmission lines and the set of \emph{directed} edges $\mcE \subseteq [N] \times [N]$ encode the cause-effect directions. Specifically, a directed edge from node $i$ to node $j\neq i$ states that if we fix the states of all nodes in the network and only alter the state of node $i$, this alteration induces a change in the state of node $j$.
In cases when a change in a node induces a change in a group of nodes, we denote the set of parents and children of node $i$ in graph $\mcG$ by $\pa(i)\subseteq [N]$ and $\ch(i)\subseteq [N]$, respectively. According to this model, when node $i$ is \textcolor{black}{failed}, it can potentially impose anomalies only on the nodes in $\ch(i)$. Conversely, each node $i$ can become anomalous only if one of the nodes in $\pa(i)$ is \textcolor{black}{failed}. We say that node $i$ is a \emph{direct} cause of node $j$ if $j \in \ch(i)$ and an \emph{indirect} cause of node $j$ if there is a direct \emph{path} from $i$ to $j$ but $j \notin \ch(i)$.
If at least two nodes in the network can mutually impose an anomaly on one another, the graph $\mcG$ becomes \emph{cyclic}, and otherwise, it is free of cycles, i.e.,~\emph{acyclic}. 

\begin{figure}[t]
    \centering
    \includegraphics[width=0.75\textwidth]{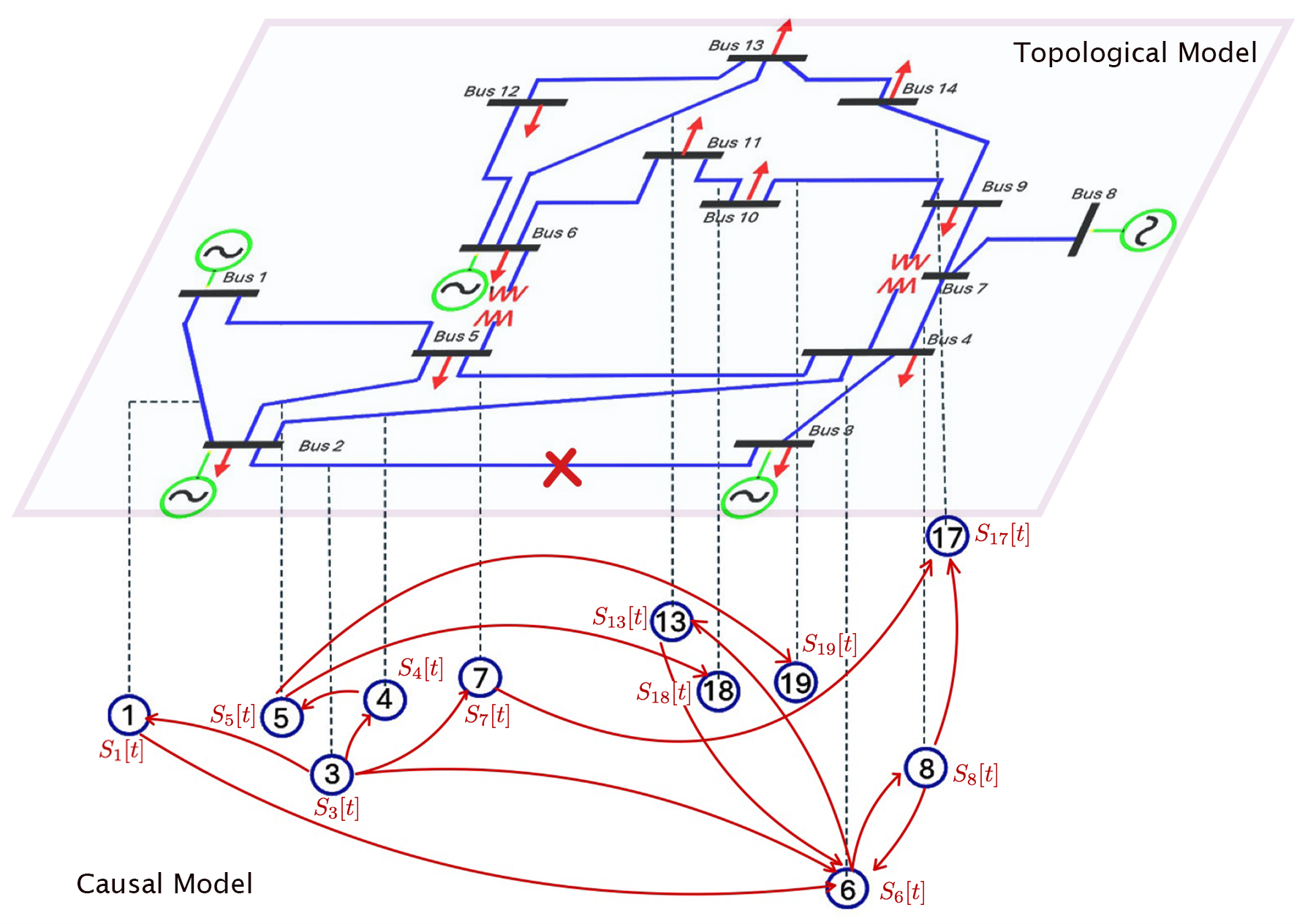}
    \caption{Causal graph for the IEEE $14$-bus system for $\mcU_{1} = \{3\}$.}
    \label{fig:main_diagram_paper}
\end{figure}

We emphasize the distinction between the graph $\mcG$ and the topological graph, the latter of which defines the physical connections among transmission lines. In contrast, graph $\mcG$ captures the latent relationships between transmission lines and characterizes how these relationships influence the propagation of \textcolor{black}{failures} in the network. Figure~\ref{fig:main_diagram_paper} illustrates the differences between the topological and learned causal model $\mcG$ (derived via Algorithm \ref{alg:Algorithm_1}, the details of which we discuss in Section~\ref{sec: Learning a Causal Graph from Observational Data}) for the IEEE $14$-bus system when line 3 (between buses 2 and 3) is an initiating \textcolor{black}{failure}.
 
\vspace{.1 in}
\noindent{\bf 2.\;Structural Causal Models:} The directed graph $\mcG$ specifies the instantaneous cause-effect relationships between components. However, it cannot specify the extent of these relationships. SCMs provide a mathematical framework for quantifying these cause-effect relationships where the state of node $i \in [N]$ at time $t$ is determined by its parent's state, denoted by $\mathbf{S}_{\text{pa}(i)}[t] \dff \{ {S_j[t] : j\in \text{pa}(i)} \}$, according to
\begin{align}
    \label{eq:SCM}
    S_i[t] \; = \; h_i(\bS_{\pa(i)}[t] ) +  \epsilon_i\ ,
\end{align} where $h_i$ represents the function describing how the states of node $i$'s parents influence node $i$ and $\epsilon_i$ accounts for the impact of unobserved random causes on node $i$. Note that the function $h_i$ may involve a linear or non-linear transformation of the parent's state. In this paper, we focus on~\emph{linear} SCMs with additive noise~\cite{ICA_LiNGAM, DAG_NOTEARS} to capture the cause-effect relationship between transmission line states, i.e., 
\begin{align}
    \label{eq:Lin_SCM}
    \bS[t] \dff \bB \cdot \bS[t] + \bm{\epsilon}\ ,
\end{align} 
where $\bB \in \mathbb{R}^{N \times N}$ is a latent causal interaction matrix. Specifically, $\bB_{ij} \neq 0$ if and only if $i \in \ch(j)$ in $\mcG$, and $\bm{\epsilon} \in \mathbb{R}^{N}$ denotes the random vector of unobserved causes, such that each random variable $\epsilon_i$ is independent of the other.

\vspace{.1 in}
\noindent{\bf 3.\;Interventions:} 
In causal inference, a key tool that enables discerning cause-effect relationships is \emph{interventions}, which enforce a change in the data mechanism of a node. This enables tracing how controlling the behavior of one node affects those of the rest. Upon applying interventions, the data collected will be referred to as interventional data, distinguishing them from \emph{observational data} gathered without any intervention. A widely used class of intervention is the $do$ interventions. Performing a $do$ intervention on node $i$ decouples its dependence from its parents, specified in~\eqref{eq:SCM}, and sets its value to a constant~\cite{causal_book}, i.e., $S_i[t] = c$ where $c\in \R$ is the intervened value. When only node $i$ is intervened, it affects only the relationship between node $i$ and its parents, while the interactions among other nodes remain unchanged.

\subsection{Learning a Causal Graph from Observational Data}
\label{sec: Learning a Causal Graph from Observational Data}

Accurate prediction of the latent anomaly states crucially relies on both the causal graph $\mathcal{G}$ and the causal interaction matrix $\mathbf{B}$, as they collectively govern the propagation of \textcolor{black}{failures} in the network. As we will discuss in Section~\ref{sec: Predicting cascading failures via Causal Reasoning}, the mutual impact of anomalies between the transmission lines can be best captured by a~\emph{cyclic} directed graph. While causal discovery in acyclic graphs is studied extensively, counterpart studies on the cyclic models are far more sparse, due to their more complex cause-effect structures. 
An effective approach to causal discovery in directed cyclic graphs is the cyclic-$\lingam$ algorithm~\cite{Cyclic_LiNGAM}, which infers the causal graph $\mcG$ and the interaction matrix $\bB$ from only \textcolor{black}{continuous-valued~\emph{observational} data. This method assumes \emph{causal sufficiency}, which ensures that all common causes of the observed variables are included in the SCMs. This assumption is significant because the presence of unobserved confounders can bias the estimated causal effects and lead to incorrect predictions of the cascading sequences.} Next, we provide the key steps involved in cyclic-$\lingam$ to learn $\bB$. We denote the historical dataset utilized for learning $\bB$ using cyclic-$\lingam$ by $\bar \bS_{\rm train}$. In Section~\ref{sec: Creating an Observational Dataset} we will discuss how such training data is created.

\vspace{.1 in}
\noindent {\bf Step~1 (Noise Decomposition):} The causal discovery objective is to learn matrix $\bB$, which governs the relationships among different entries of $\bS[t]$ specified by~\eqref{eq:Lin_SCM}. The first step of the process is noise decomposition via independent component analysis (ICA)~\cite{ICA_first}. For this purpose, we define $\bA\dff\mathbb{I}_N-\bB$. When $\bA$ is invertible, $\bar \bS_{\rm train}$ can be alternatively represented as a linear combination of $\bepsilon$
\begin{align}
\label{eq:ica}
    \bar \bS_{\rm train} = (\mathbb{I}_N-\bB)^{-1} \cdot  \bm{\epsilon} = \bA^{-1} \cdot \bm{\epsilon}\ .
\end{align} 
ICA algorithms find an invertible linear transformation $\bW_{\ica}\in \R^{N\times N}$ of the data $\bar \bS_{\rm train}$ with the objective that renders the noise distribution $\bepsilon$ to be maximally non-Gaussian and independent. Following this objective, the obtained matrix $\bW_{\ica}$ can be identified up to scaling and permutation of $\bA^{-1}$ as long as the observed data distribution $\bar{\bS}_{\rm train}$ is a linear and invertible mixture of independent noise components $\epsilon_{i}$. To eliminate relationships characterized by low strength within $\bW_{\ica}$, we employ the sparse-ICA~\cite{sparse_ica} algorithm to estimate $\bW_{\ica}$ from $\bar{\bS}_{\rm train}$, thereby rendering the latent cause-effect relationships encoded in $\bB$ sparse. Since $\bW_{\ica}$ is equivalent to $\bA$ only up to proper scaling and permutation, in the next two steps we apply proper permutation (Step 2) and scaling (Step 3) on $\bW_{\ica}$ to transform it to a high-fidelity estimate of $\bA$.

\vspace{.1 in}
\noindent {\bf Step~2 (Finding the best Assignment):} When the underlying graph $\mcG$ is acyclic, an optimal assignment $\bW_{\ica}$ is unique~\cite{ICA_LiNGAM}. However, when $\mcG$ is cyclic, the optimal assignment may not be unique. Due to the permutation indeterminacy of ICA in Step~1, the rows of $\bW_{\ica}$ are in random order. To find the correct correspondence between the independent components $\epsilon_{i}$ and the data $\bar{\bS}_{\rm train}$, we need to permute the rows of $\bW_{\ica}$ to obtain the correct correspondence. Let $\bpi \in \Pi$ denote a row permutation matrix such that the new matrix $\bpi\cdot\bW_{\ica}$ has the rows of $\bW_{\ica}$ rearranged, where $|\Pi| = N!$. We leverage Mutry's assignment~\cite{Murty} algorithm to construct a $\bpi^{*}$ as follows to identify the best choice for $\bW^{*}_{\ica}\dff \bpi^{*} \cdot \bW_{\ica}$.
\begin{align}
\label{eq:best_assignment}
    \bpi^{*} = \argmin_{\bpi \in \Pi} \quad \sum_{i}\;\frac{1}{|(\bpi \cdot \bW_{\ica})[i, i]|}\ .
\end{align}

\begin{algorithm}[t]
\caption{Cyclic-$\lingam$}
\label{alg:Algorithm_1}
\begin{algorithmic}
    \State \textbf{Input Dataset:} $\bar \bS_{\rm train}$ $m \times n$ data matrix $(n \ll m)$
    \State  \textbf{Step 1:} Obtain $\bW_{\ica}$ using sparse-ICA~\cite{sparse_ica}
    \State  \textbf{Step 2:} Find the best assignments from~\eqref{eq:best_assignment}, obtain $\bW^{*}_{\ica}\dff \bpi^{*} \cdot \bW_{\ica}$
    \State \textbf{Step 3:} Re-normalize each row of $\bW^{*}_{\ica}$~\eqref{eq:scaling_ica}
    \State  \textbf{Step 4:} Compute $\bB = \mathbb{I}_N - \bW^{*}_{\ica}$
    \State \noindent \Return $\bB$
\end{algorithmic}
\end{algorithm}

\noindent {\bf Step~3 (Scaling):} 
Scaling indeterminacy of ICA is solved by assuming all independent components $\epsilon_i$ to have unit variance and scaling $\bW^{*}_{\ica}$ appropriately. Contrary to this assumption, in cyclic-$\lingam$, independent components $\epsilon_i$ can have arbitrary variance values, thereby retaining the variance of $\epsilon_i$, the diagonal elements of $\bW^{*}_{\ica}[i,i] = 1\;\forall i \in [N]$~\cite{ICA_LiNGAM}. This necessitates to re-normalize the rows of $\bW^{*}_{\ica}$ so that all the diagonal elements equal 1, i.e.,

\begin{align}
\label{eq:scaling_ica}
     \bW^{*}_{\ica}[i,j] = \frac{ \bW^{*}_{\ica}[i,j] } {\bW^{*}_{\ica}[i,i]} \ , \quad \forall j\in[N]\ .
\end{align}

\vspace{.1 in}
\noindent {\bf Step~4 (Obtaining $\bB$):} Finally, the set of causal interaction matrices $\bB$ can be obtained as $\bB = \mathbb{I}_N - \bW^{*}_{\ica}$. Algorithm \ref{alg:Algorithm_1} summarizes the key steps involved in learning $\bB$.

\textcolor{black}{Under the assumptions that SCMs are linear and can be represented by a directed graph; the error terms $\epsilon_i$ have non-zero variance; the training samples $\bar{\bS}_{\rm train}$ are independently and identically distributed (with at most one $\epsilon_i$ being Gaussian); and all the noise terms $\epsilon_i$ are jointly independent, the Cyclic-$\lingam$ algorithm provides theoretical guarantees \cite{Cyclic_LiNGAM} 
for learning the SCMs that reflect the true distribution of the observational data $\bar{\bS}_{\rm train}$ in the large-sample limit.} 

\section{Predicting Cascading \textcolor{black}{Failures} via Causal Reasoning}
\label{sec: Predicting cascading failures via Causal Reasoning}

In this section, we present~{data-driven} algorithms for solving the two objectives for predicting cascades specified by $\mcP_1$ and $\mcP_2$. These algorithms are agnostic to the system's topology and model parameters. First, we discuss the process of collecting an observational dataset $\bar \bS_{\rm train}$, which serves as the basis for learning the matrix $\bB$. This learned matrix will subsequently guide our approach to inferring the sequences of \textcolor{black}{failures}.

\subsection{Creating an Observational Dataset $\bar \bS_{\rm train}$}
\label{sec: Creating an Observational Dataset}
 
Learning a matrix $\bB$ that can capture all the intricate cause-effect dependencies among anomalies in various lines critically depends on rich training data $\bar \bS_{\rm train}$ that adequately captures \textcolor{black}{dynamics of the system}. To create a comprehensive dataset \textcolor{black}{aligned with the inherent assumptions of cyclic-$\lingam$,} we monitor the variations in the anomaly index $S_{i}[t]$ in~\eqref{eq: anomaly_metric} for each line $i \in [N]$. These variations are guided by the stochastic nature of electricity consumption, where fluctuations in loads $j \in [L]$ are typically observed. To capture various anomalous scenarios in $\bar \bS_{\rm train}$, we monitor these variations for an initiating anomaly in~\emph{each} line $k \in [N]$.
In particular, we vary each load $j \in [L]$ to generate a time-series load profile $\mcL$ consisting of $|\mcL|$ steps that capture smooth fluctuations in loads. Subsequently, for each line \textcolor{black}{failure} $k \in [N]$ and loading condition $l \in \mcL$, we calculate the power flow in line $i\in [N]$ under \textcolor{black}{failure} condition $P^{l_{k}}_i[1]$ and flows $P^{l}_{i}[0]$ under normal condition in order to compute the anomaly index $S^{l_{k}}_i$ for each line $i$ via~\eqref{eq: anomaly_metric}. This results in a total of \textcolor{black}{causally sufficient} $|\mcL| \cdot L$ training data samples $\bS^{l_{k}}$ collectively denoted by
\begin{align}
\label{eq: training_data}
    \bar{\bS}_{\rm train}^k \dff \{ \bS^{l_{k}} \;:\;  l\in \mcL\} \ ,
\end{align} for each line \textcolor{black}{failure} $k \in [N]$, rendering an extensive dataset $\bar{\bS}_{\rm train} \dff \bigcup_{k=1}^{N} \bar{\bS}_{\rm train}^k$ \textcolor{black}{for avoiding any unobserved confounding effects.}
Each sub-dataset $\bar{\bS}_{\rm train}^k$ \textcolor{black}{is continuous-valued and} captures the extent to which an anomaly in a line $k \in [N]$ influences the anomalous states of other lines. The data generating mechanism $\bar{\bS}_{\rm train}^k$ is cyclic because each stage in a cascade is a phase-space transition from one stable equilibrium to another~\cite{cont_cas}, where the anomaly states $S^{l_{k}}_i$ and $S^{l_{k}}_j~\forall i, j \in [N]$ interact mutually and reach a corresponding steady state. More details for creating the dataset $\bar{\bS}_{\rm train}$ are provided in Section~\ref{sec: learning B}.

\subsection{Modeling Dynamics of a Cascade Using Intervention}
\label{sec: Modeling Dynamics of a Cascade}

Each occurrence of \textcolor{black}{a failure} alters the system dynamics, necessitating adjustments to the learned matrix $\bB$. Although learning a causal graph from empirical observations at each stage $m$ of the cascade offerse a potential solution, the process becomes combinatorially complex as both the system size $N$ and the horizon $M$ increase, necessitating scenario-based simulations. To address this issue, we employ~\emph{interventions} to alter the learned matrix $\bB$ appropriately at each stage of the cascade. This subsection elaborates the approach for representing a specific sequence of \textcolor{black}{failures} through a corresponding sequence of interventions.

Corresponding to a given~\emph{ordered} sequence of existing \textcolor{black}{line failures} $\mcV$ in the system, we define an updated graph $\mcG(\mcV)$ representing the latent causal structure that captures the cause-effect dynamics in the system that has undergone a sequence of \textcolor{black}{failures} in $\mcV$. We denote the corresponding SCM matrix by $\bB(\mcV)$. We treat the initiating \textcolor{black}{failures} and the subsequent ones differently because the initiating \textcolor{black}{failure} occurs due to exogenous random disturbances while the subsequent ones occur as a consequence of the initiating \textcolor{black}{failure}. 

\vspace{.1 in}
\noindent {\bf Initiating \textcolor{black}{Failures}:} Corresponding to each possible initiating \textcolor{black}{line failure}, we specify one latent causal structure $\mcG(\mcU)$ and the associated causal matrix $\bB(\mcU)$ for $\mcU=\{i\}$ as follows. When the initiating \textcolor{black}{line failure} is $i\in [N]$, we set $\mcU=\{i\}$. Next, we learn the set of causal matrices $\mcB$
\begin{align}
    \label{eq:causal_matrix_set}
    \mcB \dff \{\bB(\mcU): \mcU = \{i\}, i \in [N]\}\ ,
\end{align} using the corresponding observational dataset $\bar{\bS}_{\rm train}^i$ in~\eqref{eq: training_data} by leveraging the Cyclic-$\lingam$ procedure presented in Algorithm~\ref{alg:Algorithm_1}. Each causal matrix $\bB(\mcU) \in \mcB$ captures the extent to which an initiating \textcolor{black}{line failure} $\mcU$ influences the anomaly states of its children $j\in \ch(\mcU_m)$. 

\vspace{.1 in}
\noindent {\bf Cascade of \textcolor{black}{Failures}:} Consider stage $m\in[M]$ of the cascade, up to which the ordered set of lines  $\mcN_m \triangleq \langle \mcU_1, \dots, \mcU_{m} \rangle$ have \textcolor{black}{failed}. We leverage the matrix $\bB(\mcU_{m}) \in \mcB$ learned for the initiating \textcolor{black}{failure} $\mcU_m$ and consider intervention mechanisms to find the updated $\bB(\mcN_m)$ as follows. 
\begin{align}
\label{eq:update_rule}
    [\bB(\mcN_m)]_{i,j} = \left\{
    \begin{matrix}
 0 &  i\in \mcN_{m-1} \\  
         & \\ 
        [\bB(\mcU_m)]_{i,j} & i\notin \mcN_{m-1}
\end{matrix}\right. \ .
\end{align} At stage $m$, the lines included in the set $ \mcN_{m-1}$ are already \textcolor{black}{in failure condition} and cannot be affected further by their parents. Based on this, the update rule in~\eqref{eq:update_rule} is specified to ensure that only the causal relationships associated with the most recent \textcolor{black}{failure} $\mcU_m$ are considered.

\subsection{$\cpath$ Algorithm: Real-Time Anomaly Prediction}
\label{subsec:path_analysis}

\begin{algorithm}[t]
\caption{$\cpath$ Algorithm}
\label{alg:Algorithm_2}
	\begin{algorithmic}
        \Function{C-Path}{$\mcN_m$, $\bB(\mcN_m)$, \textcolor{black}{$\kappa \%$}} \funclabel{alg:one} 
        \State $ \mcQ \leftarrow [N] \setminus \mcN_{m} $
        \State $\mcX \leftarrow \langle \rangle$
        \For{$j \in  \mcQ$}
            \State Compute $D_m(j)$ using~\eqref{eq:causal_effect}~\Comment{\textcolor{red}{total causal effect}}\
            \State $\mcX$.append( $|D_m(j)|$ )
        \EndFor
        \State $\mcQ \gets$ Sort $\mcQ$ based on decreasing $\mcX$
        \State $\hat{\mcU}_{m + 1} \leftarrow \mcQ[1:\textcolor{black}{\kappa \%}]$
        \State \noindent \Return {$\hat{\mcU}_{m + 1}$}
        \EndFunction
	\end{algorithmic}
\end{algorithm}

In $\mcP_1$, our objective is the real-time prediction of the next \textcolor{black}{failure} when a cascade has begun, i.e., in stage $m$ of the cascade, predicting the transmission lines that will \textcolor{black}{fail} at stage $m+1$. This prediction relies on the hidden dependencies among nodes extracted from the dynamic causal matrices $\bB(\mcN_m)$, which is found via a sequence of interventions according to the rule in~\eqref{eq:update_rule}.

At stage $m$, given the most recent \textcolor{black}{line failure} $\mcU_m$ and given the causal matrix $\bB(\mcN_m)$, we specify how to predict a set of lines causally influenced by $\mcU_m$, which we denote by $\hat{\mcU}_{m+1}$. The influence of \textcolor{black}{failed} $\mcU_m$ can propagate through the paths involving the descendants of $\mcU_m$ in graph $\mcG(\mcN_m)$. 
To obtain the aggregate causal effects along all the direct paths from $i \in \mcU_m$ to $j \in [N]\backslash\mcN_m$, we leverage causal path analysis~\cite{LLC}, discussed next.

\vspace{0.1 in}

\noindent {\bf Causal path analysis:} We provide an analysis substantiating that the \textcolor{black}{$\cpath$} algorithm provides a solution for $\mcP_1$. The cascades of \textcolor{black}{failures} propagate both locally and non-locally~\cite{IG_1}. At stage $m$, the direct path from $\mcU_m$ to $j \in [N]\backslash\mcN_m$ is denoted by $\mcU_m \curly j$. We denote the set of all such direct paths in $\mcG(\mcN_m)$ 
by  
$\mcK_{m,j}$. The contribution of each path $\ell \in \mcK_{m,j}$ is specified by the product of the edge coefficients on this path, denoted by
\begin{align}
    X_{\ell} = \prod_{(n \rightarrow k) \in \ell} [\bB(\mcN_m)]_{n,k} \ .
\end{align}
By summing the contributions of all paths $\ell \in \mcK_{m,j}$, and normalizing over all $j \in [N]\setminus \mcN_m$, the total influence is
\begin{align}
  \label{eq:causal_effect}
    D_m(j) = \frac{|\sum_{\ell \in \mcK_{m,j}} X_{\ell}|}{\sum_{j \in [N] \setminus \mcN_m} | \sum_{\ell \in \mcK_{m,j}} X_{\ell}| }\ ,
\end{align}
which quantifies the cumulative effect of the anomaly state, $S_{\mcU_m}[m]$, on line $j$. The metric $D_m(j)$ acts as a proxy for the probabilities $\P(j|\bar{\bB}_m, \bar{\bS}_m)$ in~\eqref{eq:P1} extent to which \textcolor{black}{a failure} at $\mcU_m$ can be amplified as it propagates through the network. 
Finally, we select $\hat{\mcU}_{m+1}$ as the set of lines with the top \textcolor{black}{$\kappa \%$} associated terms in $\{D_m(j): j \in [N]\backslash\mcN_m\}$
\begin{align}
\label{eq:top_kappa}
 \hat \mcU_{m+1} \triangleq \argmax_{\mcU\subseteq [N] \backslash \mcN_m, |\mcU| \leq \textcolor{black}{\lceil N \times \kappa \% \rceil }}\sum_{j\in\mcU} D_m(j) \ .
\end{align} The $\cpath$ algorithm is summarized in Algorithm~\ref{alg:Algorithm_2}.


\subsection{Critical Cascade Identification (CCI)}

In problem $\mcP_2$, we are interested in obtaining the top $d$ most costly sequences out of all possible cascades. Similar to solving $\mcP_1$, we adopt causal reasoning to identify these sequences. Specifically, we recursively implement the $\cpath$ algorithm and generate a set of predicted sequences. This process has two stages, specified next.

\noindent{\bf Exploration:} For each initiating \textcolor{black}{failure} $i \in [N]$ in stage $m = 1$, the CCI algorithm starts by predicting the \textcolor{black}{$ \kappa \% $} most probable lines to be \textcolor{black}{failed} at $m = 2$ using the $\cpath$ Algorithm. This set is specified by $\hat{\mcU}_2$. Considering one line can \textcolor{black}{fail} at each stage of the cascade, for each line $j \in \hat{\mcU}_2$, we can further obtain the set of top \textcolor{black}{$\kappa \% $} lines that are most likely to \textcolor{black}{fail} at $m =3$. This set, in turn, is denoted by $\hat{\mcU}_3$. This recursive process extends to all stages $m \in [M]$ to explore the sequences until it reaches the maximum depth $M$. This process obtains $N \times \textcolor{black}{\lceil N \times \kappa \% \rceil}^{M-1}$ sequences as the candidates for being a solution to $\mcP_2$. The recursive exploration steps are outlined in Algorithm~\ref{alg:Algorithm_3}.

\begin{algorithm}[t]
\caption{CCI Explore}
\label{alg:Algorithm_3}
    \begin{algorithmic}
\Function{GenSeq}{$\mcN_m, \textcolor{black}{\kappa\%}, M$}
\funclabel{alg:gen_helper} 
    \If{$|\mcN_m| = M$}
        \State \Return $\{ \mcN_m \}$
    \EndIf	
    \State $\mcC_{i} \leftarrow \emptyset$
    \State Construct $\bB(\mcN_m)$ using~\eqref{eq:update_rule}
    \State $\hat{\mcU}_{m+1} \gets \Call{C-Path}{\mcN_m, \bB(\mcN_m), \textcolor{black}{\kappa \%}}$~\Comment{\textcolor{red}{Algorithm~\ref{alg:Algorithm_2}}}
    \For{$\mcU\in\hat{\mcU}_{m+1}$}
        \State $\mcN_m$.append($\mcU$)
        \State $\mcC_{i} \leftarrow \mcC_{i} \bigcup \Call{GenSeq}{\mcN_m, \textcolor{black}{\kappa \%}, M}$
    \EndFor
    \State \Return $\mcC_{i}$
    \EndFunction 

    \State $\mcC \gets \emptyset$~\Comment{\textcolor{red}{Main Loop}}
     \For{$i \in [N]$}
         \State $\mcN_m \leftarrow \langle i \rangle $
        \State{$\mcC \leftarrow \mcC\bigcup \Call{GenSeq}{\mcN_m, \textcolor{black}{\kappa \%}, M}$}
    \EndFor
    \State \Return $\mcC$~\Comment{\textcolor{red}{Set of Predicted Sequences}}
\end{algorithmic}
\end{algorithm}

\begin{algorithm}[t]
\caption{CCI Algorithm}
\label{alg:Algorithm_4}
    \begin{algorithmic}
    \State $S \gets [\hspace{0.1cm}]$
    \State $\mcC \leftarrow \text{CCI Explore via Algorithm~\ref{alg:Algorithm_3}}$
    \For{$\mcV_i \in \mcC$}
    \State Compute $S_i[m]$ using~\eqref{eq: anomaly_metric} via AC power flow simulations 
    \State Compute $\tv(\mcV_i)$ using~\eqref{eq:total_vulnerability}
    \State $S$.append($\tv(\mcV_i)$)
    \EndFor
    \State $\mcC \gets$ Sorted $\mcC$ based on decreasing $S$
    \State $\bar{\mcV}_d \gets \mcC[1:d]$
    \State \Return  $\bar{\mcV}_d$   
\end{algorithmic}
\end{algorithm}

\noindent{\bf Critical Cascade Identification:} Upon obtaining a set of candidates $\mcC$, we calculate the cost corresponding to each of the sequences specified by~\eqref{eq:total_vulnerability} and identify the top $d$ most costly sequences denoted by $\bar{\mcV}_d$. We emphasize that the exploration process significantly reduces the search space for such sequences. Moreover, utilizing the $\cpath$ algorithm to obtain the sequences provides us with the critical transmission lines in each stage $m$, which helps us explore the local and non-local behavior of \textcolor{black}{a failure} sequence. Based on this, the steps for selecting the top $d$ most costly sequences are summarized in Algorithm~\ref{alg:Algorithm_4} and are specified to ensure efficient exploration in the causally guided search space.

\section{Case Studies: Cascading Line Outages}
\label{sec: Case Studies: Anomalies as Line Outages}

In this section, we focus on line outage cascades, an important class of cascading failures, and assess the performance of our proposed causal inference framework for solving problems $\mcP_1$ and $\mcP_2$ on the \textcolor{black}{IEEE $14$-bus, IEEE $39$-bus, and IEEE $118$-bus} systems. We employ the AC power-flow model in MATPOWER to generate the observational dataset $\bar{\bS}_{\rm train}$ required for the various inferential decisions involved. To formalize line outages, we adopt a dichotomous \textcolor{black}{failure} state model $\tilde S_i[t] \in \{0,1\}$, where $\tilde S_i[t]=0$ indicates normal operation (i.e., $P_i[t] < P_i^{\sf max}$), while $\tilde S_i[t]=1$ represents a line outage (i.e., $P_i[t] \geq P_i^{\sf max}$). Accordingly, \textcolor{black}{a failure} sequence $\mcV$ in~\eqref{eq:sequence} signifies the progression of cascading line outages.

\subsection{Model Training and Evaluation}
\label{sec: Model Training and Evaluation}

\subsubsection{Generating Observational Data and Model Training}
\label{sec: learning B}

Accurate causal predictions critically depend on generating proper training data. We begin by defining a baseline loading condition, referred to as the $\baseload$. Using this baseline, we generate a smooth time-series load profile consisting of $|\mcL| =$ 20,000 steps, where in each step, the loads are uniformly sampled within a range of $90$\% to $110$\% of the $\baseload$, and a smoothing kernel is applied to ensure realistic load variations. Subsequently, corresponding to each loading condition $l \in \mcL$ and each initiating anomalous line $k \in [N]$, we compute the anomaly index $S^{l_{k}}_i$ in~\eqref{eq: anomaly_metric} for each line $i \in [N]$ using AC power flows and accordingly, construct the dataset $\bar{\bS}^k_{\rm train}$ in~\eqref{eq: training_data}. Note that we only consider initiating outages that do not induce islands in the network. Subsequently, we learn the set of causal matrices $\mcB$ in~\eqref{eq:causal_matrix_set} associated with each dataset $\bar{\bS}^k_{\rm train}$ using Algorithm~\ref{alg:Algorithm_1}. These matrices are then subjected to interventions, following the procedure outlined in rule~\eqref{eq:update_rule}, to predict \textcolor{black}{line outages}.


\subsubsection{Model Evaluation}
\label{sec: Model Evaluation}

In our causal inference framework, the step following training is inference. Consequently, assessing the quality of the inferential predictions made by the learned models $\mcB$ is essential for validating their reliability. This validation process necessitates the generation of ground truth data, which provides a benchmark for accurately measuring the predictive accuracy of the models. \textcolor{black}{We implement a cascading failure simulator (CFS) based on the AC power flow module in MATPOWER. Specifically, due to time constraints, the operators often lack the ability to optimally re-dispatch generation and instead rely primarily on automatic generation control (AGC) to quickly reestablish power balance.} We next outline this process through the following six steps, in which we describe how we generate the set of all~\emph{dependent} cascading line outage sequences $\mcZ$ occurring over a maximum of $M$ stages using our \textcolor{black}{CFS}:

\begin{enumerate}[label= \textbf{CFS Step} \arabic*:, wide, labelwidth=!, labelindent=0pt, itemindent=0pt, labelsep=0.3em]
    \item Initialize AC power flow according to $\baseload$.
    \item Create an outage in line $k\in[N]$ in the network.
    \item If the network bifurcates into distinct islands, the cascade ends, and the simulation re-initializes from Step~$1$.
    \item If no islands are identified, power flows $\bP[m]$ are re-calculated to accommodate the current contingency.
    \item The CFS finds the lines $i\in [N]$ that are overloaded $P_i[m] \geq P_i^{\sf max}$. If there are no overloaded lines, the cascade ends and the simulation reinitializes from Step~$1$.
    \item If some lines are overloaded, we \emph{sequentially} remove each overloaded line one at a time (rendering multiple sequences) and the simulation re-initializes from Step~$3$.
\end{enumerate}

 We account for dependent outages until the sequence of failures ceases, which occurs under one of the following conditions: (i) the flows in each line return to within operational limits; (ii) an islanding criterion within the CFS is activated; or (iii) the cascade progresses through $M$ stages. This iterative process, from Step-$1$ to $6$, generates multiple sequences of dependent outages, each initiated by an outage in line $k$. Finally, to generate $\mcZ$, we repeat this iterative process for each potential initiating outage $k\in [N]$. When $M=4$, this procedure yields a total of $|\mcZ| = 2,213$ and $|\mcZ| = 4049$ distinct sequences of cascading failures for the IEEE $14$-bus and IEEE $39$-bus systems, respectively, using our developed CFS. \textcolor{black}{For IEEE $118$-bus system, we consider $M=3$ and generate a total of $|\mcZ| = 30,298$ distinct sequences.} Note that since we do not consider network islands, $N=19$ for the \textcolor{black}{IEEE $14$-bus, $N=35$ for the IEEE $39$-bus, and $N=177$ for the IEEE $118$-bus test cases.}

\subsection{Predicting Most Likely Cascading Line Outages ($\mcP_1$)}
\label{sec: Predicting Most Likely Cascading Line Outages}

In this section, we evaluate the $\cpath$ Algorithm~\ref{alg:Algorithm_2} using the ground truth sequences generated by the CFS, as outlined in Section~\ref{sec: Model Evaluation}. To quantify performance, we report $\Precision (\mcV)$, as defined in~\eqref{eq:cost-p2}, averaged over all the ground truth sequences $\mcV^{*} \in \mcZ$. In the context of cascading line outages, $\hat \mcU_{m}$ represents the predicted line outages by the model under evaluation, while the actual outages $\mcU_{m}$ are extracted from the ground truth sequences $\mcV^{*} \in \mcZ$. The precision metric thus quantifies the model's accuracy in predicting the most likely line outage for each of the $M-1$ subsequent stages of a cascading sequence $\mcV^{*} \in \mcZ$ initiated by line $\mcU_{1}$.

Computing the predictions $\hat \mcU_{m+1}$ in~\eqref{eq:top_kappa}  at each stage $m$, relies on computing the total causal influence from $\mcU_m$ to every other healthy line $j \in [N]$. This influence is quantified by $D_m(j)$, as specified in~\eqref{eq:causal_effect}. To calculate these causal influences, we perform a breadth-first search (BFS) within each updated graph $\mcG(\mcN_m)$ at stage $m$ to compute $D_m(j)$. Our experiments indicate that paths with length $|\ell| > 3$ contribute minimally to the magnitude of $D_m(j)$, because of the small coefficients $[\bB(\mcU_m)]_{n,k}$. Therefore, we limit our consideration to paths up to a length of $|\ell|\leq3$. Consequently, each prediction set $\hat{\mcU}_{m+1}$ is obtained by varying \textcolor{black}{$\kappa \%$}, which has two opposing impacts. On the one hand, increasing \textcolor{black}{$\kappa \%$} leads to improved precision, and on the other hand, it indicates selecting more components, thereby increasing the complexity of solving~\eqref{eq:top_kappa} due to the $\argmax$ operation. Therefore, a desired value of \textcolor{black}{$\kappa \%$} can be specified based on user-specific requirements by balancing accuracy and computational cost.

We evaluate the performance of our proposed $\cpath$ algorithm by comparing it against the decisions derived from \textcolor{black}{\emph{influence graphs} (IGs)~\cite{IG_1} and \emph{graph neural networks} (GNN) model~\cite{ML_1},} as the most relevant approaches. 

\noindent {\bf \textcolor{black}{Influence Graph Baseline:}} 
IGs are constructed under the assumption of a Markov process, where the probability of a line failing in stage $m+1$ is dependent solely on the failures in stage $m$ and remains independent of any failures in earlier stages. In this model, the IG is structured as a bipartite graph $g$ with $2N$ nodes: the first partition contains $N$ nodes 
representing the lines at stage $m$, and the second partition includes the remaining $N$ nodes representing the lines in stage $m+1$. An edge from node $i$ in the first partition to node $j$ in the second indicates the likelihood of line $j$ failing at stage $m+1$ if line $i$ failed at stage $m$. To compute the edge weights in the influence graph $g$, the learning algorithm from~\cite{IG_1} processes all stages of the training data sequences and counts the number of times that line $i$ occurs as a parent of $j$ represented by $g_c[j|i]$. The transition probabilities are then obtained by normalizing over all children outages originating from line $i$ given by $g[j|i] \triangleq \tfrac{g_c[j|i]}{\sum_{j = 1}^N g_c[j|i]}$. \textcolor{black}{During inference, at each stage $m$, we set $g[j | \mcU_m] = 0$ for all lines $j \in \mcN_{m-1}$ that are in an outage, and then identify the top $\kappa \%$ components from the re-normalized distribution $g[j | \mcU_m]$. Specifically, in each stage $m \in [M]$ of the cascade, we predict the set $\hat \mcU_{m+1}$ by the following rule
\begin{align}
\label{eq:IG CCI}
    \hat \mcU_{m+1} \triangleq \argmax_{\mcU\subseteq [N] \backslash \mcN_m, |\mcU| \leq \lceil N \times \kappa \% \rceil}\;\;g[\mcU\;|\;\mcU_{m}] \ .
\end{align} } 

\vspace{0.1 in}
\noindent \textcolor{black}{{\bf GNN Model:}
GNN is a deep learning-based flow-free approach~\cite{ML_1} that leverages the network topology, represented as a graph $\mcG_T = ([Q], [N])$, where the set of buses $[Q]$ serves as nodes and the transmission lines $[N]$ as edges. This model is used to predict line outages at each stage $m$, given an initial failure state $\tilde{\bS}[1]$ and the power injection values $\tilde{P}_i[1]$ for each bus $i \in [Q]$. The GNN model consists of four neural network blocks, with weights learned via backpropagation during the initial training phase.}

\textcolor{black}{In the first stage, a dense network transforms the bus features $\{\tilde{P}_i[1]: i \in [Q]\}$ into hidden transmission line features. The second, attention stage takes $\tilde{\bS}[1]$ as input and processes it through two dense neural networks to compute neighboring edge-to-edge and node-to-edge features. In the third, averaging stage, the output features from the previous stages are aggregated by calculating a weighted average, repeated $K$ times. Finally, in the last stage, failure probabilities $\P(i|m)$ are calculated for each component $i \in [N]$ at each stage $m \in [M]$. For inference, to integrate the GNN model into our framework, we obtain,
\begin{align}
    \hat \mcU_{m+1} \triangleq \argmax_{\mcU\subseteq [N] \backslash \mcN_m, |\mcU| \leq \lceil N \times \kappa \% \rceil } \P (\mcU | m) \ .
\end{align}} 

\vspace{0.1 in}
\noindent \textcolor{black}{{\bf Training Dataset For Baseline Models:}}
To facilitate comparisons, we trained the IG model \textcolor{black}{and the GNN model}, generating 10,000 and 20,000 training cascading sequences \textcolor{black}{$\bar{\bQ}_{\rm train}$} using the DCSIMSEP~\cite{random_chemistry} for the IEEE $14$-bus and IEEE $39$-bus systems, respectively. To tailor this model to our setting, we structured the sequences into stages, segregating outages that occur more than $\Delta t = 2$ apart, such that in each stage, $|\mcU_m|=1$. In the case where multiple transmission lines are overloaded, however, lines are not removed simultaneously. Instead, a single line is selected for removal based on a sampling process that prioritizes lines according to the normalized distribution of their fractional overload $\tfrac{P_i[m]}{P_i^{\sf max}}$ and we adhere to the stopping criteria outlined in Section~\ref{sec: Model Evaluation} for fair comparisons.


\begin{figure}[t]
    \centering
    \subfigure[Precision versus $\kappa \%$ for the IEEE $14$-bus system.]{
        \includegraphics[width=0.48\textwidth]{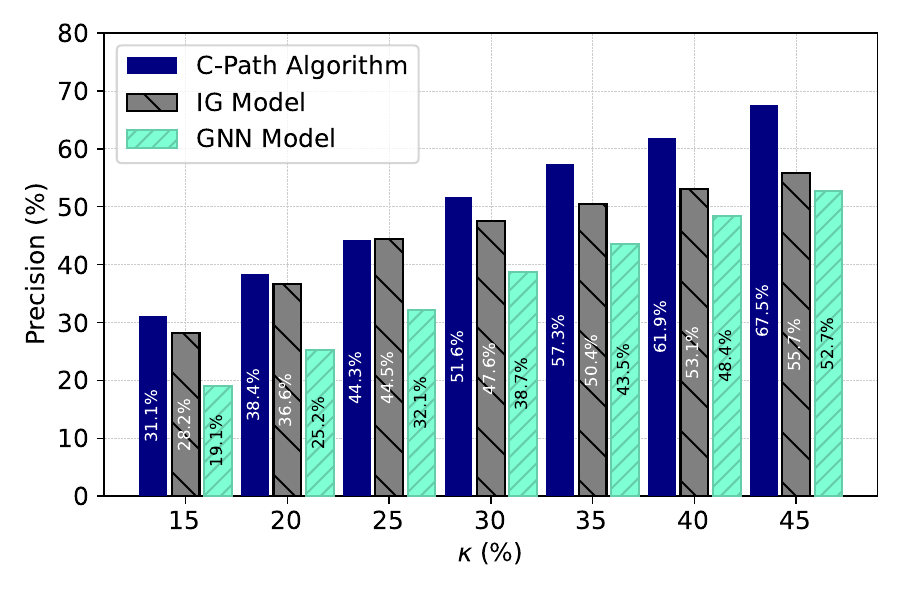}
        \label{fig:P2_14}
    }
    \hfill
    \subfigure[Precision versus $\kappa \%$ for the IEEE $39$-bus system.]{
        \includegraphics[width=0.48\textwidth]{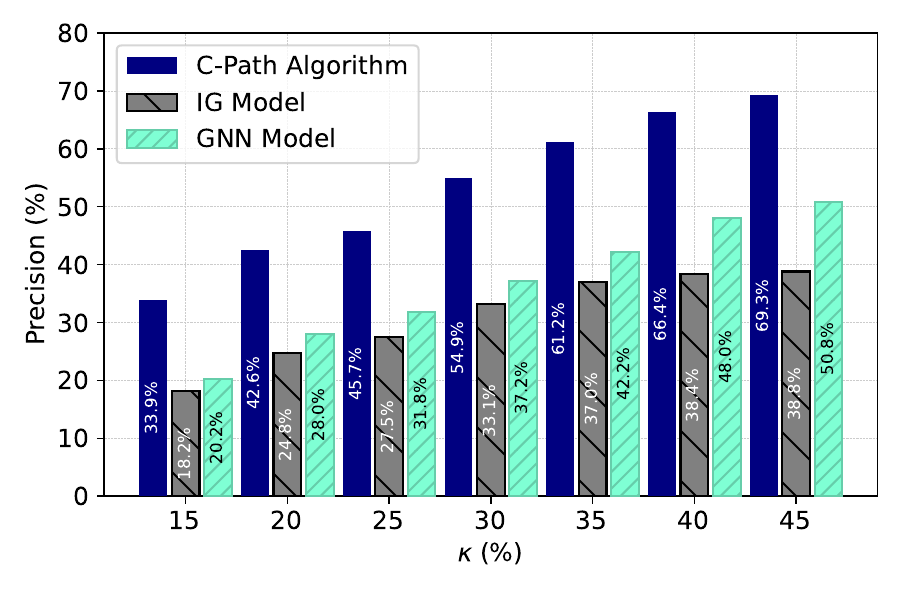}
        \label{fig:P2_39}
    }
    \caption{Precision for IEEE $14$-bus and $39$-bus systems.}
    \label{fig:comparison}
\end{figure}

Figures~\ref{fig:P2_14} and~\ref{fig:P2_39} illustrate the variations in average precision as a function of different \textcolor{black}{$\kappa \%$} values for the two test systems under evaluation. These figures also demonstrate a comparison between the accuracy of predictions obtained from our causal approach and those derived from influence graphs. We have two main observations. First, 
the $\cpath$ Algorithm can effectively capitalize on the latent dependencies learned between the line outage interactions by leveraging the total influences $D_{m}(j)$ computed via the causal matrix coefficients $[\bB(\mcN_m)]_{i,j}$. \textcolor{black}{For instance, while the prediction involves approximately $\kappa = 25\%$ of the total lines in its prediction set $\hat \mcU_{m}$, the $\cpath$ algorithm for IEEE $14$-bus, and IEEE $39$-bus achieves comparable precision accuracy of $44.3\%$, and $45.7\%$ respectively}. This demonstrates a \textcolor{black}{$82.8\%$} improvement compared to outcomes based on uniformly random selection. Secondly, the $\cpath$ Algorithm consistently outperforms IGs \textcolor{black}{and GNN models} across all \textcolor{black}{$\kappa \%$} values for \textcolor{black}{all} the systems evaluated. For example, for the IEEE $14$-bus system with \textcolor{black}{$\kappa = 45\%$}, $\cpath$ shows a $21.18\%$ improvement in precision accuracy over IGs, \textcolor{black}{and $28.08\%$ improvement in precision accuracy over GNN models.} For the IEEE $39$-bus system, \textcolor{black}{the improvement over IGs} is even more substantial \textcolor{black}{$78.60\%$} for the same \textcolor{black}{$\kappa \%$}. \textcolor{black}{The improvement over the GNN model is $36.41 \%$, which is lesser than IG but still significant compared to the causal model for $\kappa = 45\%$.}  



The significant performance discrepancy stems from the fact that the IG, \textcolor{black}{and GNN models} primarily capture the most probable cascading sequences \textcolor{black}{in the training dataset $\bar{\bQ}_{\rm train}$ which is due to the fact that in reality line outages in the system are determined by the extent of the overloads} which may not provide comprehensive insights into cascading outage predictions in~\emph{less probable} cascading sequences, as captured by the ground truth $\mcV^{*} \in \mcZ$. \textcolor{black}{We observe that the GNN model slightly outperforms the IG model for IEEE $39$-bus even when the training data is similar. This is because the GNN model uses additional features $\tilde{\bS}[1]$ and $\{\tilde{P}_i[1]: i \in [Q]\}$ aligned to the sequences in $\bar{\bQ}_{\rm train}$ while the settings in the IG model cannot accommodate these additional features. On the other hand, we observe that for the IEEE $14$-bus model, IG outperforms GNN for lower \textcolor{black}{$\kappa \%$} and performs similarly for higher \textcolor{black}{$\kappa \%$}. This can be an artifact of IG having a smaller system size compared to an over-parameterized GNN, which overfits the training dataset as we use the same network used in \cite{ML_1} for both the test cases.} 
We also note that the causal framework provides the more accurate prediction for the~\emph{most likely} cascading outages by leveraging latent causal interactions and targeted interventions agnostic to the realized scenario-based training. 




\subsection{Predicting Critical Cascading Outages ($\mcP_2$)}
\label{sec: Predicting Critical Cascading Outages}

In this section, we evaluate the predicted costly sequences $\bar{\mcV}_d$ by the CCI Algorithm (Algorithm~\ref{alg:Algorithm_4}) based on the $d$ most critical ground truth sequences $\bar \mcV_{d}^* \subseteq \mcZ$. To quantify performance, we report the $\Regret(\bar{\mcV}_d)$ in~\eqref{eq:regret} associated with each model under evaluation. Without loss of generality, we set $d=100$ and $M=4$, although any other value can be chosen. For comparisons, we construct an IG model, \textcolor{black}{and a GNN model}, based on the steps outlined in Section~\ref{sec: Predicting Most Likely Cascading Line Outages}, and obtain the $d$ sequences as follows: First, we run Algorithm~\ref{alg:Algorithm_3}~\emph{based on} the influence graph, \textcolor{black}{and the GNN model learned} from data. 
For each component $\mcU \in \hat \mcU_{m+1}$, the CCI Explore Algorithm probes a tree of possible sequences and returns the set $\mcC_{\rm IG}$, \textcolor{black}{and $\mcC_{\rm GNN}$ respectively}. Finally, Algorithm~\ref{alg:Algorithm_4} is employed to find the $d$ most costly sequences from the set $\mcC_{\rm IG}$, \textcolor{black}{and $\mcC_{\rm GNN}$} to construct $\bar{\mcV}_d$. We denote the sequences generated by our causal approach after running Algorithm~\ref{alg:Algorithm_3} by $\mcC_{\rm causal}$. We note that the number of sequences, $|\mcC|$, for both the causal and IG models can be significantly smaller than $N \times \textcolor{black}{\lceil N \times \kappa \% \rceil}^{M-1}$. This reduction can be attributed to two factors: (i) cascading sequences may terminate prematurely due to the CFS stopping criteria outlined in Section~\ref{sec: Model Evaluation}, and (ii) more critically, both the IG $g$ and causal graph $\mcG(\mcV)$ tend to be inherently sparse, resulting in fewer than \textcolor{black}{$\kappa \%$} components in the set $\hat \mcU_{m+1}$ carrying non-trivial probabilities, i.e., $|\hat \mcU_{m+1}| < \textcolor{black}{\lceil N \times \kappa \% \rceil}$.


\begin{figure}[t]
    \centering
    \subfigure[Regret$(\bar\mcV_{100})$ versus $\kappa \%$ for the IEEE $14$-bus system.]{
        \includegraphics[width=0.48\textwidth]{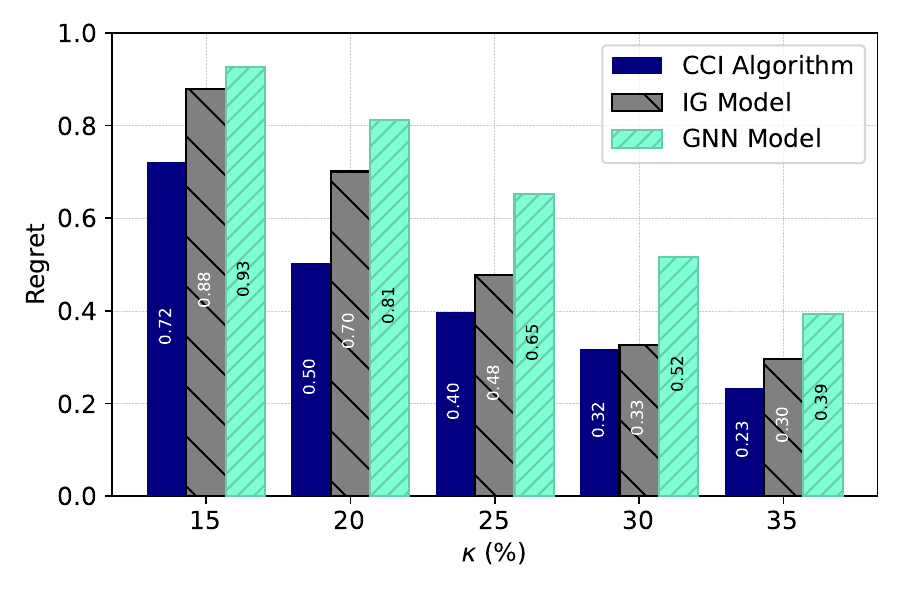}
        \label{fig:P3_ol_14}
    }
    \hfill
    \subfigure[Regret$(\bar\mcV_{100})$ versus $\kappa \%$ for the IEEE $39$-bus system.]{
        \includegraphics[width=0.48\textwidth]{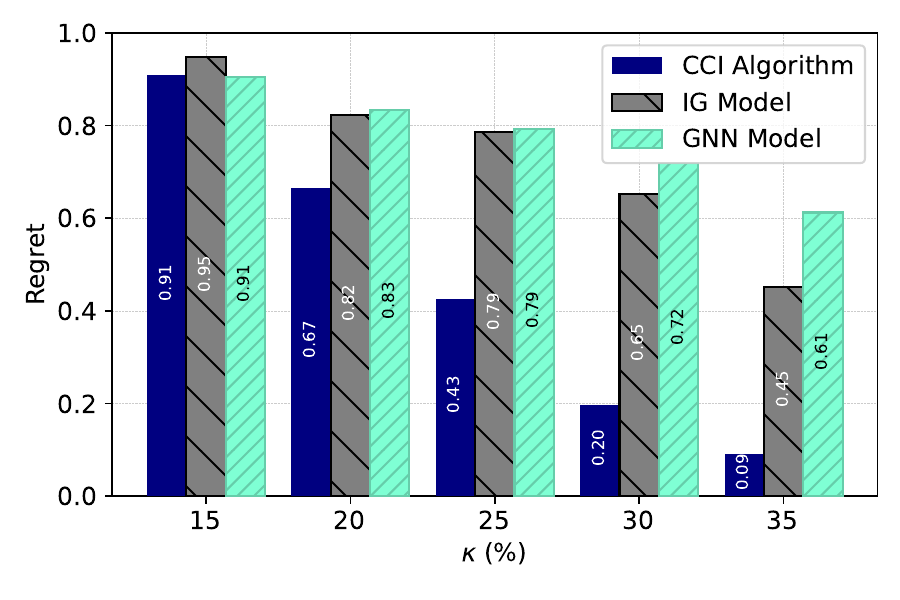}
        \label{fig:P3_ol_39}
    }
    \caption{Regret for IEEE $14$-bus and $39$-bus systems.}
    \label{fig:regret_comparison}
\end{figure}

Figures~\ref{fig:P3_ol_14} and~\ref{fig:P3_ol_39} illustrate the performance of the two test systems as a function of different \textcolor{black}{$\kappa \%$} values, while also comparing the accuracy of predictions of our causal approach to those derived from IGs, \textcolor{black}{and GNN} models. We have two main observations. First,
as \textcolor{black}{$\kappa \%$} increases, the regret associated with the CCI algorithm decreases drastically indicating that the causal approach  finds~\emph{critical} cascading outages. For instance, for the IEEE $39$-bus system at \textcolor{black}{$\kappa = 30\%$}, the CCI algorithm suffers a regret of only $0.2$ despite searching for a small fraction $\tfrac{|\mcC_{\rm causal}|}{|\mcZ|} = 8.8\%$ of total cascading sequences. The observations corroborate with that of the IEEE $14$-bus system. Secondly, 
the CCI Algorithm consistently outperforms the IG, \textcolor{black}{and GNN} model when predicting costly sequences across all \textcolor{black}{$\kappa \%$} values for both system sizes. For example, in the IEEE $14$-bus system with \textcolor{black}{$\kappa = 35\%$}, the CCI algorithm shows a \textcolor{black}{$23.33\%$} decrease in regret over IGs, while in the IEEE $39$-bus system, this improvement is even more substantial, a $45.57\%$ decrease in regret for \textcolor{black}{$\kappa= 25 \%$}. 

\begin{table}[h]
\centering
\scalebox{1}{
	\begin{tabular}{| c | c | c | c | c |} 		
		\hline
		\thead{\textcolor{black}{$\kappa \%$}}               
		& \thead{$N \times \textcolor{black}{\lceil N \times \kappa \% \rceil}^{M-1}$}   
		& \thead{$|\mcC_{\rm IG}|$} &  \textcolor{black}{\thead{$|\mcC_{\rm GNN}|$}} 
		& \thead{$|\mcC_{\rm Causal}|$} \\ 
		\hline
		\textcolor{black}{15}   & 7,560    & 35    & \textcolor{black}{35}   & 62   \\ 
		 
		\textcolor{black}{20}   & 17,920   & 50  & \textcolor{black}{49} & 145     \\  
	    \textcolor{black}{25}   & 25,515   & 62  & \textcolor{black}{56}   & 166   \\
	    
        \textcolor{black}{30}  & 46,585   & 92 & \textcolor{black}{65}  & 349   \\ 
        \textcolor{black}{35} &  \textcolor{black}{76,895}     &  \textcolor{black}{127}  &  \textcolor{black}{85} & \textcolor{black}{458} \\
		\hline
	\end{tabular}%
	}
	\caption{\textcolor{black}{Critical cascading sequences discovered for IEEE $39$-bus.}}
	\label{table: numSequences}
\end{table}

This significant performance discrepancy is due to the fact that the learned IG model is relatively sparser than the causal model, as illustrated from Table~\ref{table: numSequences}. Specifically, in this table, the second \textcolor{black}{column} lists the theoretical maximum number of sequences for each initiating outage $N$ for various \textcolor{black}{$\kappa \%$} values. The third and fourth columns compare the number of critical sequences identified by the IG and causal models, respectively. As \textcolor{black}{$\kappa \%$} increases, the growth in critical sequences identified by the IG model plateaus, which suggests a sparser IG model due to fewer components in the set $\hat \mcU_{m+1}$. \textcolor{black}{However, the GNN model is denser as it computes the probability of an outage at each stage $m \in [M]$, yet it does not perform as well as the causal model due to the tendency of deep learning models to overfit the training data $\bar{\bQ}_{\rm train}$. While the training data primarily captures the most probable but often less costly cascading sequences, using lower \textcolor{black}{$\kappa \%$} does not effectively account for $\bar{\mcV}_{100}$. Conversely, at higher \textcolor{black}{$\kappa \%$} values, the GNN model does not exhibit the plateau effect, and at $\kappa = 100\%$, it can successfully identify all $\bar{\mcV}_{100}$ sequences.}

\begin{figure}[t]
    \centering
    \includegraphics[width=0.6\textwidth]{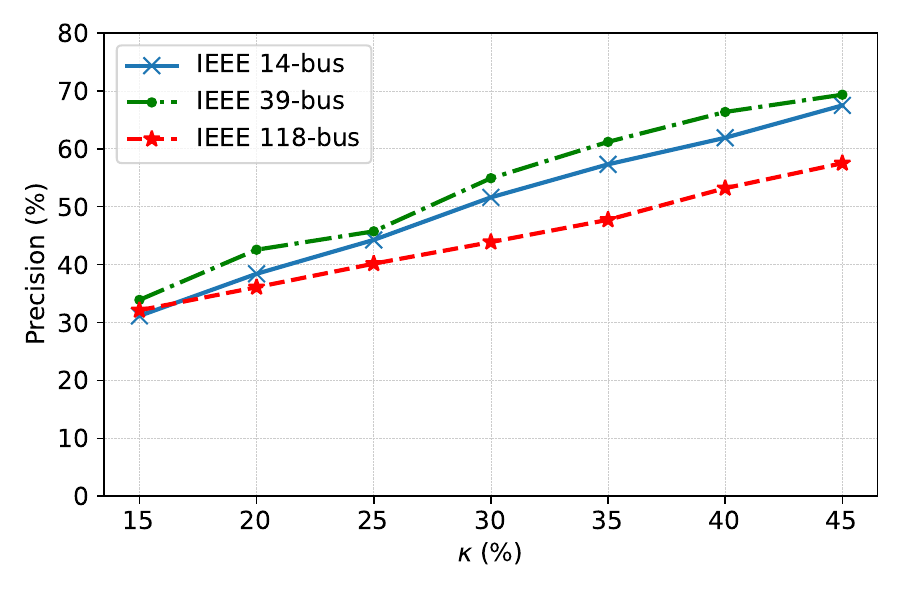}
    \captionsetup{skip=-4pt}
    \caption{\textcolor{black}{Precision versus \textcolor{black}{$\kappa \%$} for IEEE $14$-bus, $39$-bus, and $118$-bus systems.}}
    \label{fig:P2_scalability}
\end{figure}

\textcolor{black}{\subsection{Scalability}
In this section, we provide a comparative study comprising IEEE $14$-bus, $39$-bus, and $118$-bus test cases for demonstrating the scalability of the $\cpath$ algorithm. 
For the larger IEEE $118$-bus system, the training becomes expensive as we have to learn $\mcB$, which is a set of $177$ matrices. However, once the model is trained, we can obtain the prediction by choosing the appropriate \textcolor{black}{$\kappa \%$}, which leads to a trade-off between accuracy and complexity.}

\textcolor{black}{Figure~\ref{fig:P2_scalability} illustrates the comparison of three test cases as a function of different \textcolor{black}{$\kappa \%$}. We can observe the consistent performance of the $\cpath$ algorithm for increasing \textcolor{black}{$\kappa \%$}. For instance, when the prediction set $\hat{\mcU}_m$ includes approximately $\kappa = 25\%$ of the total lines, the $\cpath$ algorithm achieves comparable precision accuracies of $44.3\%$, $45.7\%$, and $40.2\%$ for the IEEE $14$-bus, $39$-bus, and $118$-bus systems, respectively. Although there is an approximate $12\%$ decrease in precision for the IEEE $118$-bus test case, the algorithm still demonstrates a $60.8\%$ improvement compared to the outcomes of uniformly random selection for the same $\kappa \%$.}
\begin{figure}[t]
    \centering
    \includegraphics[width=0.6\textwidth]{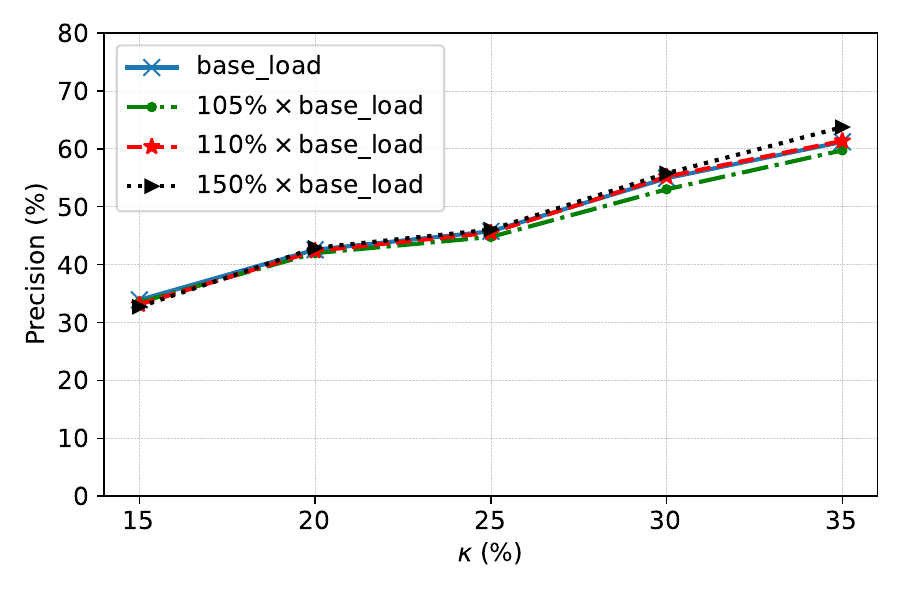}
    \captionsetup{skip=-4pt}
    \caption{\textcolor{black}{Precision versus \textcolor{black}{$\kappa \%$} for different loading condition for IEEE $39$-bus system}}
    \label{fig:robustness}
\end{figure}

\textcolor{black}{\subsection{Robustness}
Any variation in loading conditions alters the ground truth against which the model's performance needs to be evaluated. In this subsection, we evaluate the robustness of our algorithm to load variations.  
To systematically explore the impact of these variations, we test the model under increased loading conditions of $105\% \times \baseload$, $110\% \times \baseload$, and $150\% \times \baseload$, resulting in 4,333, 4,625, and 36,674 ground truth sequences ($|\mcZ|$), respectively. The results of these tests are illustrated in Figure~\ref{fig:robustness}.}

\textcolor{black}{
 A key observation is that with an increase in the loading condition, we can see a significant increase in the ground truth cascades $|\mcZ|$. Despite these changes, the $\cpath$ algorithm demonstrates consistent performance across all \textcolor{black}{$\kappa \%$} thresholds and loading conditions. In practical scenarios, loading conditions typically vary uniformly across all buses, except in cases where a fault occurs at a load bus. Consequently, power redistribution tends to follow a uniform pattern. This uniformity allows a single learned causal graph to effectively predict the most likely cascading failures across different loading scenarios.}

\subsection{Worst-Case Sequences}

Besides the naturally occurring failures, power systems are also vulnerable to failures imposed by external forces (e.g., cyber-attacks). Causal frameworks can also be adapted to address such failures, which is a major difference from IG-based approaches, which work based on naturally occurring overloads and line outages. 

To accommodate cascading failures due to external forces, finding the most costly sequence may require searching through all worst-case sequences of length $M=4$, rendering the problem combinatorially complex. This process generates 79,628 and 1,042,674 distinct cascading failure sequences for the IEEE  $14$-bus and IEEE $39$-bus systems, respectively. Identifying the worst possible case necessitates evaluating all possible sequences to obtain the $d$  most costly sequences. Some sequences may not occur in the set of dependent overloading sequences; if they occur, they can incur huge costs. For instance, in the IEEE $39$-bus model, the sequence $\langle 10,9,12,7\rangle$ cannot occur due to only naturally occurring overloads. As a result, this sequence remains undetectable by the IG approach, while the CCI Algorithm can detect it. Table~\ref{tab2} represents the most costly sequences identified by the causal and IG models for $d=1$. It is observed that for the IEEE $14$-bus system, the sequences identified by the CCI and IG algorithms share the same lines (but in different orders) and comparable costs ($71.33$ versus $71.24$). However, for the IEEE $39$-bus system, the line outages identified are entirely different, and the one identified by the CCI algorithm has a notably higher cost ($37.15$ versus $27.5$). 
\begin{figure}[t]
    \centering
    \includegraphics[width=0.6\textwidth]{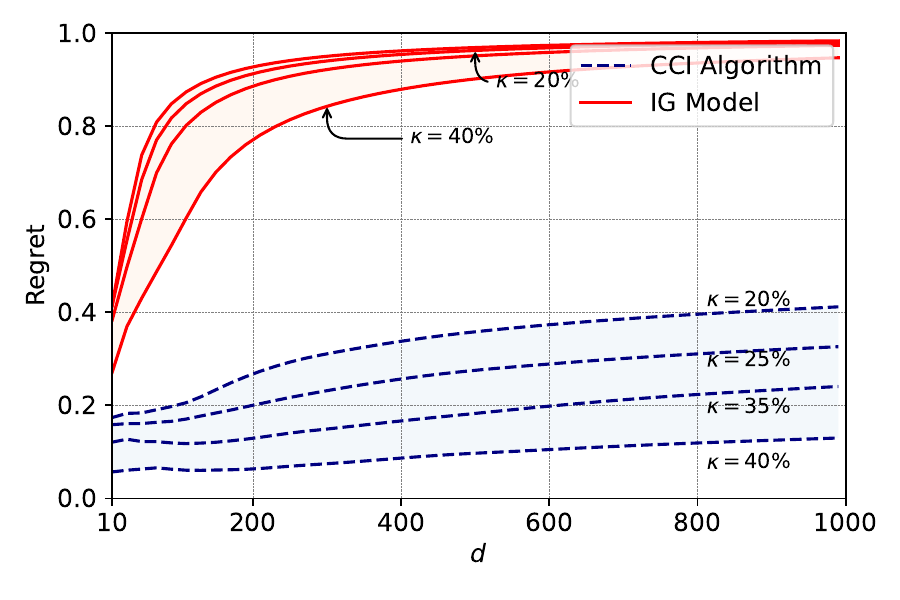}
    \captionsetup{skip=-4pt}
    \caption{\textcolor{black}{Regret for varying $d$ mostly costly sequences (IEEE $39$-bus)}}
    \label{fig:P3_39}
\end{figure}

\begin{table}[h]
  \centering
  \begin{tblr}{|Q[c,1.5cm]|Q[c,1.9cm]|Q[c,1.4cm]|Q[c,1.9cm]|Q[c,1.4cm]|Q[c,0.5cm]|}
    \hline
   System & CCI   & $\tv(\mcV_{1})$ & IG & $\tv(\mcV_{1})$ & \textcolor{black}{$\kappa \%$} \\
   \hline
   IEEE  $14$   & $\langle 7,4,3,11\rangle$ & $71.33$ & $\langle 3,7,4,11 \rangle$ & 71.24 & \textcolor{black}{20} \\
   \hline
   IEEE $39$ & $\langle 10,9,12,7\rangle$  & $37.15$ & $\langle 2, 29, 8, 12\rangle$ & $27.5$  & \textcolor{black}{35}\\
   \hline
  \end{tblr}
  \caption{\textcolor{black}{Costliest predicted worst-case sequence.}}
  \label{tab2}
\end{table}

We expand the above analysis and in Figure~\ref{fig:P3_39} present a comprehensive summary of the results for a wider range of \textcolor{black}{$\kappa \%$} and $d$. While both CCI and IG have a small gap for lower values of $d$, the gap sharply increases as $d$ increases. Specifically, for $d=50$, the IG method's performance quickly rises to within $80\%$ of its maximum value. \textcolor{black}{In contrast, the CCI algorithm shows a steady improvement towards} its maximum regret, which is about $40\%$ of the regret of the IG approach.

Finally, we note that despite the notable improvements in the performance, the proposed causal methods have low computational complexity since predictions essentially involve simple matrix multiplications and additions. In Table~\ref{tab:computation_time}, we compare the computational ties of CCI and IG algorithms.
\begin{table}[h]
    \centering
    \begin{tabular}{|c|c||c|c|}
    \hline
      \textcolor{black}{$\kappa \%$}   & CCI Complexity (s) & IG Complexity (s)  \\
      \hline
      \textcolor{black}{5}   & 0.37 & 0.29 \\
      \textcolor{black}{20}  & 14.5  & 6.10\\
      \textcolor{black}{45}  & 125.13 &  43.32\\
      \hline
    \end{tabular}
    \caption{\textcolor{black}{Computational time comparison for $\mcP_{2}$ (IEEE $14$-bus).}}
    \label{tab:computation_time}
\end{table}

\section{Concluding Remarks}

In this paper, we addressed the challenge of predicting and identifying the most likely and most costly cascading \textcolor{black}{failures} within power system networks using exclusively observational data. Our proposed framework, built upon advancements in causal inference, leverages a causal graph to model component interactions, accounting for the local and non-local anomaly propagation in networks. Empirical assessments conducted on the \textcolor{black}{IEEE $14$-bus, $39$-bus, and $118$-bus} systems validate the efficacy of our framework in predicting critical cascades within the system. By exclusively utilizing observational data and without relying on multiple ground-truth scenarios for training, our causal modeling approach offers better efficiency and performance compared to the existing state-of-the-art baseline.

\textcolor{black}{Our primary focus has been on line outages. However, this framework can be further expanded to accommodate a broader range of transmission line anomalies that subsume line outages as extreme cases. 
Furthermore, we have also been focusing on outages (anomalies) that occur one at-a-time. This is, however, for the simplicity of describing algorithm details and is not a fundamental assumption. In the same way that single-line outages have been modeled via single-node interventions, we can model multiple concurrent failures as multiple concurrent interventions to recover the causal model of the system after multiple failures from the anomaly-free systems. Even though our focus has been on one failure at-a-time, our model can be generalized to account for multiple failures, too. 
}

\textcolor{black}{
Finally, we remark that model robustness is a critical aspect of causal inference, particularly in the presence of data inaccuracies and missing data. Our framework employs the Cyclic-$\lingam$ algorithm for learning the causal model, which has inherent robustness against non-Gaussian noise in observational data. This algorithm is designed to effectively filter out non-Gaussian disruptions, making it resilient to data inaccuracies that extend beyond the typical Gaussian noise models. However, handling missing data remains a limitation of the Cyclic-$\lingam$ algorithm, as it does not account for unobserved confounders, which can affect the accuracy of causal relationships. This remains an open question to be addressed in causal inference and its application to power systems when there is missing data.}

\bibliographystyle{IEEEtran}
\bibliography{cascade}

\end{document}